\documentclass[thmsa]{article}
\usepackage{sw20lart}

\input{tcilatex}
\begin{document}

\title{Entropic Dynamics\thanks{%
Presented at MaxEnt 2001, the 21th International Workshop on Bayesian
Inference and Maximum Entropy Methods (August 4-9, 2001, Baltimore, MD, USA).%
}}
\author{Ariel Caticha \\
{\small Department of Physics, University at Albany-SUNY, }\\
{\small Albany, NY 12222, USA.\thanks{%
E-mail: Ariel.Caticha@albany.edu}}}
\date{}
\maketitle

\begin{abstract}
I explore the possibility that the laws of physics might be laws of
inference rather than laws of nature. What sort of dynamics can one derive
from well-established rules of inference? Specifically, I ask: Given
relevant information codified in the initial and the final states, what
trajectory is the system expected to follow? The answer follows from a
principle of inference, the principle of maximum entropy, and not from a
principle of physics. The entropic dynamics derived this way exhibits some
remarkable formal similarities with other generally covariant theories such
as general relativity.
\end{abstract}

\section{Introduction}

The study of changes in the natural world, dynamics, is divided among
several distinct disciplines. Thermodynamics, for example, considers changes
between special states, the so-called states of equilibrium, and addresses
the question of which final states can be reached from any given initial
state. Mechanics studies the changes we call motion, chemistry deals with
chemical reactions, quantum mechanics with transitions between quantum
states, and the list goes on.

In all of these examples we want to predict or explain the observed changes
on the basis of information that is codified in a variety of ways into what
we call the states. In some cases the final state can be predicted with
certainty, in others the information available is incomplete and we can, at
best, only assign probabilities.

The theory of thermodynamics holds a very special place among all these
forms of dynamics. With the development of statistical mechanics by Maxwell,
Boltzmann, Gibbs and others, and eventually culminating in the work of
Jaynes \cite{Jaynes57}, thermodynamics became the first clear example of a
fundamental physical theory that could be derived from general principles of
probable inference. The entire theory follows from a clear idea of the
subject matter, that is, an appropriate choice of which states one is
talking about, plus well-known principles of inference \cite{footnote1},
namely, consistency, objectivity, universality and honesty. These principles
are sufficiently constraining that they lead to a unique set of rules for
processing information: these are the rules of probability theory \cite
{Cox46} and the method of maximum entropy \cite{Jaynes57}\cite
{ShoreJohnson80}.

There are strong indications that a second example of a dynamics that can be
deduced from principles of inference is afforded by quantum mechanics \cite
{Caticha98}. Many features of the theory, traditionally considered as
postulates, follow from the correct identification of the subject matter
plus general principles of inference. Briefly, the goal of quantum mechanics
is not to predict the behavior of microscopic particles, but rather to
predict the outcomes of experiments performed with certain idealized setups.
Thus, the subject of quantum theory is not just the particles, but rather
the experimental setups. The variables that encode the information relevant
for prediction are the amplitudes or wave functions assigned to the setups.
These ingredients plus a requirement of consistency (namely, that if there
are two ways to compute an amplitude, the two results should agree)
supplemented by entropic arguments are sufficient to derive most of the
standard formalism including Hilbert spaces, a time evolution that is linear
and unitary, and the Born probability rule.

If quantum mechanics, deemed by many to be \emph{the }fundamental theory,
can be derived in this way, then it is possible, perhaps even likely, that
other forms of dynamics might ultimately reflect laws of inference rather
than laws of nature. Should this turn out to be the case, then the
fundamental equations of change, or motion, or evolution as the case might
be, would follow from probabilistic and entropic arguments and the discovery
of new dynamical laws would be reduced to the discovery of what is the
necessary information for carrying out correct inferences. Unfortunately,
this search for the right variables has always been and remains to this day
the major stumbling block in the understanding of new phenomena.

The purpose of this paper is to explore this possible connection between the
fundamental laws of physics and the theory of probable inference: Can
dynamics be derived from inference? Rather than starting with a known
dynamical theory and attempting to derive it, I proceed in the opposite
direction and ask: What sort of dynamics can one derive from
well-established rules of inference?

In section 2 I establish the notation, define the space of states, and
briefly review how the introduction of a natural quantitative measure of the
change involved in going from one state to another turns the space of states
into a metric space \cite{Caticha01a}. (Such metric structures have been
found useful in statistical inference, where the subject is known as
Information Geometry \cite{Amari85}, and in physics, to study both
equilibrium \cite{Weinhold75} and nonequilibrium thermodynamics \cite
{Balian86}.)

Typically, once the kinematics appropriate to a certain motion has been
selected, one proceeds to define the dynamics by additional postulates. This
is precisely the option I want to avoid: in the dynamics developed here
there are no such postulates. The equations of motion follow from an
assumption about what information is relevant and sufficient to predict the
motion.

In a previous paper \cite{Caticha01a} I tackled a similar problem. There I
answered the question:

\begin{description}
\item[Q1:]  Given the initial state and that the system evolves to other
states, what trajectory is the system expected to follow?
\end{description}

\noindent This question implicitly assumes that there is a trajectory and
that information about the initial state is sufficient to determine it. The
dynamical law follows from the application of a principle of inference, the
method of maximum entropy (ME), to the only information available, the
initial state and the recognition that motion occurred. Nothing else. The
resulting `entropic' dynamics is very simple: the system moves continuously
and \emph{irreversibly} along the entropy gradient.

Thus, the honest, correct answer to the inference problem posed by question
Q1 has been given, but the equally important question `Will the system in
fact follow the expected trajectory?' remained unanswered. Whether the
actual trajectory is the expected one depends on whether the information
encoded in the initial state happened to be sufficient for prediction.
Indeed, for many systems, including those for which the dynamics is \emph{%
reversible}, more information is needed.

In section 3 we answer the question:

\begin{description}
\item[Q2:]  Given the initial and the final states, what trajectory is the
system expected to follow?
\end{description}

\noindent Again, the question implicitly assumes that there is a trajectory,
that in moving from one state to another the system will pass through a
continuous set of intermediate states. And again, the equation of motion is
obtained from a principle of inference, the principle of maximum entropy,
and not from a principle of physics. (For a brief account of the ME method
in a form that is convenient for our current purpose see \cite{Caticha01b}.)
The resulting `entropic' dynamics also turns out to be simple: the system
moves along a geodesic in the space of states. This is simple but not
trivial: the geometry of the space of states is curved and possibly quite
complicated.

Important features of this entropic dynamics are explored in section 4. We
show that there are some remarkable formal similarities with the theory of
general relativity (GR). For example, just as in GR there is no reference to
an external physical time. The only clock available is provided by the
system itself. It turns out that there is a natural choice for an
`intrinsic' or `proper' time. It is a derived, statistical time defined and
measured by the change itself. Intrinsic time is quantified change. This
entropic dynamics can be derived from a Jacobi-type principle of least
action, which we explore both in Lagrangian and Hamiltonian form. Just as in
GR there is invariance under arbitrary reparametrizations -- a form of
general covariance -- and the entropic dynamics is an example of what is
called a constrained dynamics.

\section{Quantifying change: geometry}

In this section we briefly review how to quantify the notion of change (for
more details see \cite{Caticha01a}). The idea is simple: since the larger
the change involved in going from one state to another, the easier it is to
distinguish between them, we claim that change can be measured by
distinguishability. Next, using the ME method one assigns a probability
distribution to each state. This transforms the problem of distinguishing
between two states into the problem of distinguishing between the
corresponding probability distributions. The solution is well-known: the
extent to which one distribution can be distinguished from another is given
by the distance between them as measured by the Fisher-Rao information
metric \cite{Fisher25}\cite{Rao45}\cite{Amari85}. Thus, change is measured
by distinguishability which is measured by distance.

Let the microstates of a physical system be labelled by $x$, and let $m(x)dx$
be the number of microstates in the range $dx$. We assume that a state of
the system (\emph{i.e.}, a macrostate) is defined by the expected values $%
A^{{}\alpha }$ of some $n_A$ appropriately chosen variables $a^{{}\alpha
}(x) $ ($\alpha =1,2,\ldots ,n_A$), 
\begin{equation}
\left\langle a^{{}\alpha }\right\rangle =\int dx\,p(x)a^{{}\alpha
}(x)=A^{{}\alpha }\,.  \label{Aalpha}
\end{equation}
A crucial assumption is that the selected variables codify all the
information relevant to answering the particular questions in which we
happen to be interested. This is a point that we have made before but must
be emphasized again: there is no systematic procedure to choose the right
variables. At present the selection of relevant variables is made on the
basis of intuition guided by experiment; it is essentially a matter of trial
and error. The variables should include those that can be controlled or
observed experimentally, but there are cases where others must also be
included. The success of equilibrium thermodynamics, for example, derives
from the fact that a few variables are sufficient to describe a static
situation, and being few, these variables are easy to identify. In fluid
dynamics, on the other hand, the selection is more dificult. One must
include many more variables, such as the local densities of particles,
momentum, and energy, that are neither controlled nor usually observed.

The states form an $n_A$-dimensional manifold with coordinates given by the
numerical values $A^{{}\alpha }$. To each state we can associate a
probability distribution $p(x|A)$. The distribution that best reflects the
prior information contained in $m(x)$ updated by the information $%
A^{{}\alpha }$ is obtained by maximizing the entropy 
\begin{equation}
S[p:m]=-\int \,dx\,p(x)\log \frac{p(x)}{m(x)}.  \label{S[p]}
\end{equation}
subject to the constraints (\ref{Aalpha}). The result is 
\begin{equation}
p(x|A)=\frac 1Z\,m(x)\,e^{-\lambda _{{}\alpha }a^{{}\alpha }(x)},
\label{pzero}
\end{equation}
where the partition function $Z$ and the Lagrange multipliers $\lambda
_{{}\alpha }$ are given by 
\begin{equation}
Z(\lambda )=\int dx\,m(x)\,e^{-\lambda _{{}\alpha }a^{{}\alpha }(x)}\quad 
\text{and}\quad -\frac{\partial \log Z}{\partial \lambda _{{}\alpha }}%
=A^{{}\alpha }\,.  \label{Z and lambda}
\end{equation}

Next, we argue that the change involved in going from state $A$ to the state 
$A+dA$ can be measured by the extent to which the two distributions can be
distinguished. As discussed in \cite{Amari85}, except for an overall
multiplicative constant, the measure of distinguishability we seek is given
by the `distance' $d\ell $ between $p(x|A)$ and $p(x|A+dA)$, 
\begin{equation}
d\ell ^2=g_{\alpha \beta }\,dA^{{}\alpha }dA^{{}\beta }\,\,,
\end{equation}
where 
\begin{equation}
g_{\alpha \beta }\,=\int dx\,p(x|A)\,\frac{\partial \log p(x|A)}{\partial
A^{{}\alpha }}\,\frac{\partial \log p(x|A)}{\partial A^{{}\beta }}\,\,
\end{equation}
is the Fisher-Rao metric \cite{Fisher25}\cite{Rao45}. It turns out that this
metric is unique: it is the only Riemannian metric that adequately reflects
the fact that the states $A$ are not `structureless points', but happen to
be probability distributions.

To summarize: the very act of assigning a probability distribution $p(x|A)$
to each state $A$, automatically provides the space of states with a metric
structure.

\section{Dynamics and intrinsic time}

Given the initial and the final states, what trajectory is the system
expected to follow? The key to answering this question lies in the implicit
assumption that there exists a trajectory or, in other words, that large
changes are the result of a continuous succession of very many small
changes. Thus, the difficult problem of studying large changes is reduced to
the much simpler problem of studying small changes.

Let us therefore focus on small changes and assume that the change in going
from the initial state $A_i$ to the final state $A_f=A_i+\Delta A$ is small
enough that the distance $\Delta \ell $ between them is given by 
\begin{equation}
\Delta \ell ^2=g_{\alpha \beta }\,\Delta A^{{}\alpha }\Delta A^{{}\beta }\,.
\end{equation}
To find which states are expected to lie on the trajectory between $A_i$ and 
$A_f$ we reason as follows. In going from the initial to the final state the
system must pass through a halfway point, that is, a state $A$ that is
equidistant from $A_i$ and $A_f$ (see fig.1a). The question is which halfway
state should we choose? An answer to this question would clearly determine
the trajectory: first find the halfway point, and use it to determine
`quarter of the way' points, and so on.

Next we notice that there is nothing special about halfway states. We could
equally well have argued that in going from the initial to the final state
the system must first traverse a third of the way, that is, it must pass
through a state that is twice as distant from $A_f$ as it is from $A_i$. In
general, we can assert that the system must pass through intermediate states 
$A_{{}\omega }$ such that, having already moved a distance $d\ell $ away
from the initial $A_i$, there remains a distance $\omega d\ell $ to be
covered to reach the final $A_f$. Halfway states have $\omega =1$, `third of
the way' states have $\omega =2$, and so on (see fig.1b).

\FRAME{fthFU}{4.6553in}{3.0978in}{0pt}{\Qcb{The geometry of dynamics: in
going from the initial$\ A_i$ to the final state $A_f$ the system must pass
through states $A$ (dashed line) that are equidistant between them (a), and
it must also pass through states $A_{{}\omega }$ (dashed circle) that are $%
\omega $ times as far from $A_f$ as they are from $A_i$ (b).}}{}{ed-fig.eps}{%
\special{language "Scientific Word";type "GRAPHIC";maintain-aspect-ratio
TRUE;display "USEDEF";valid_file "F";width 4.6553in;height 3.0978in;depth
0pt;original-width 558.0625pt;original-height 430.625pt;cropleft "0";croptop
"1";cropright "1";cropbottom "0.1398";filename
'C: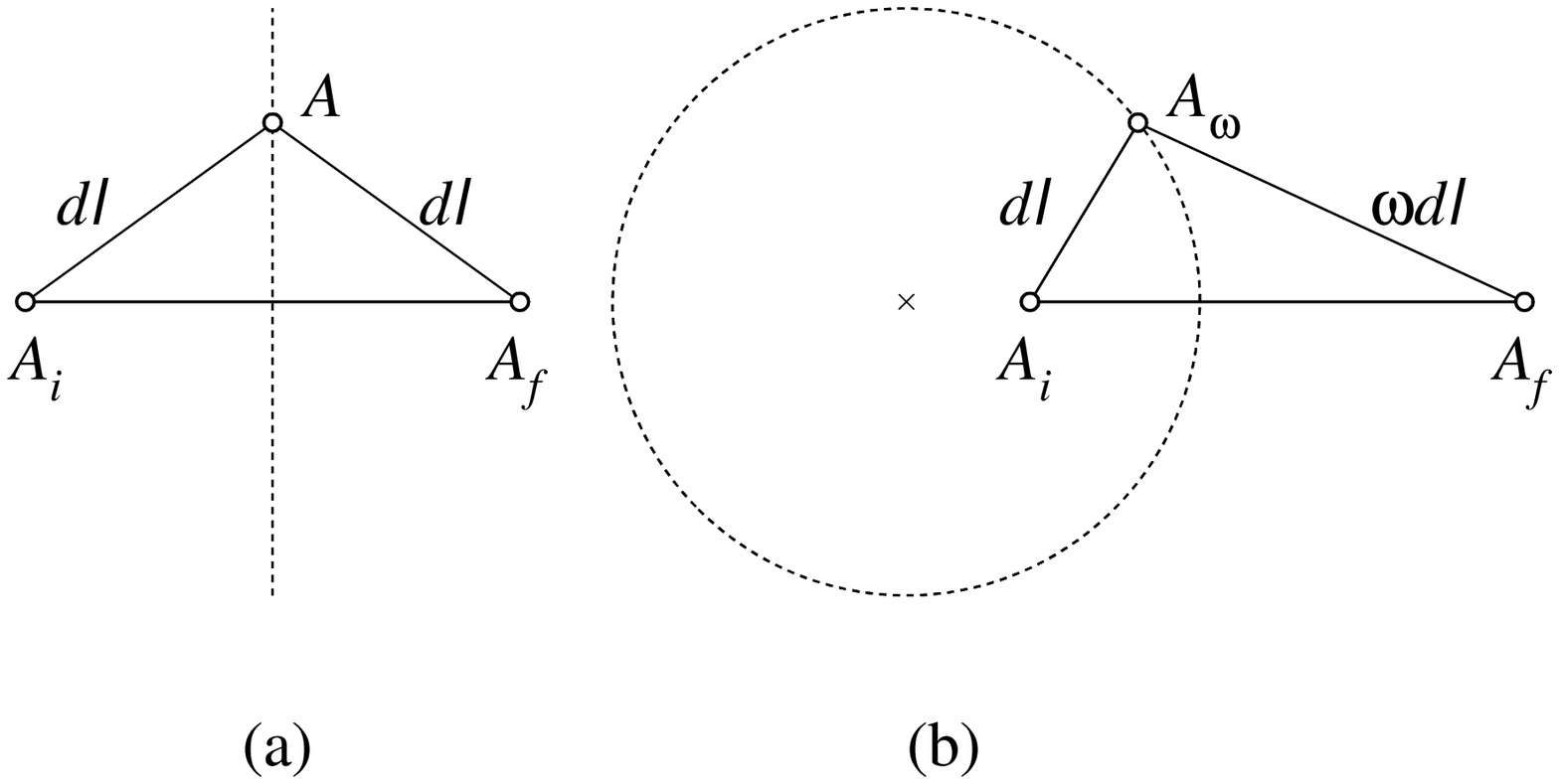';file-properties "XNPEU";}}

It appears that each different value of $\omega $ provides a different
criterion to select the trajectory. If there is a trajectory and there are
several ways to determine it, consistency demands that all these ways should
agree: in the end we must verify that the selected trajectory is independent
of $\omega $ or else we have a problem.

Our basic dynamical question Q2 can be rephrased as follows:

\begin{description}
\item[Q2$^{\prime }$]  The system is initially in state $p(x|A_i)$ and we
are given the new information that the system has moved to one of the
neighboring states in the family $p(x|A_{{}\omega })$. Which $%
p(x|A_{{}\omega })$ do we select?
\end{description}

\noindent Phrased in this way it is clear that this is precisely the kind of
problem to be tackled using the ME method. Recall \cite{Caticha01b}: The ME
method is a method for processing information and change our minds. It
allows us to go from an old set of beliefs, described by the prior
probability distribution, to a new set of beliefs, described by the
posterior distribution, when the information available is just a
specification of the family of distributions from which the posterior must
be selected \cite{footnote2}. In the more traditional applications of the
method this family of posteriors is constrained or defined by the known
expected values of some relevant variables, but this is not necessary, the
constraints need not be linear functionals. Here the constraints are defined
geometrically.

An important question that should arise whenever one contemplates using the
ME method is which entropy should one maximize. Since we want to select a
distribution $p(x|A)$ the entropies to be considered must be of the form 
\begin{equation}
S[p:q]=-\int \,dx\,p(x|A)\log \frac{p(x|A)}{q(x)}.
\end{equation}
This is the entropy of $p(x|A)$ relative to the \emph{prior} $q(x)$. The
interpretation of $q(x)$ as the prior follows from the logic behind the ME
method itself. Recall \cite{Caticha01b}: In the absence of new information
there is no reason to change one's mind. When there are no constraints the
selected posterior distribution should coincide with the prior distribution.
Since the distribution $p$ that maximizes $S[p:q]$ subject to no constraints
is $p\propto q$, we must set $q(x)$ equal to the prior.

Coming back to our dynamical problem, suppose we know that the system is
initially in state $p(x|A_i)$ and we are not given the information that the
system moved, then we have no reason to suspect that any change has
occurred. Therefore the prior $q(x)$ should be chosen so that the
maximization of $S[p:q]$ subject to no constraints yields the posterior $%
p=p(x|A_i)$. The correct choice is $q(x)=p(x|A_i)$.

Now we are ready to tackle the question Q2$^{\prime }$: the answer is
obtained by maximizing the entropy 
\begin{equation}
S[A:A_i]=-\int \,dx\,p(x|A)\log \frac{p(x|A)}{p(x|A_i)},
\end{equation}
subject to the constraint $A=A_{{}\omega }$. This presents no problems. It
is convenient to write $A_{{}\omega }=A_i+dA$ and $A_f=A_i+\Delta A$ so that 
$S[A_{{}\omega }:A_i]$ simplifies to 
\begin{equation}
S[A_i+dA:A_i]=-\frac 12g_{\alpha \beta }\,dA^{{}\alpha }dA^{{}\beta },
\end{equation}
and the distances $d\ell _i$ and $d\ell _f$ from $A_{{}\omega }$ to $A_i$
and $A_f$ are given by 
\begin{equation}
d\ell _i^2=\,g_{\alpha \beta }\,dA^{{}\alpha }dA^{{}\beta }\quad \text{and}%
\quad d\ell _f^2=g_{\alpha \beta }\,(\Delta A^{{}\alpha }-dA^{{}\alpha
})(\Delta A^{{}\beta }-dA^{{}\beta }).  \label{distances}
\end{equation}
Then, to maximize $S[A_i+dA:A_i]$ under variations of $dA$ subject to the
constraint 
\begin{equation}
\omega d\ell _i=d\ell _f\,,  \label{w constr}
\end{equation}
introduce a Lagrange multiplier $\lambda $, 
\begin{equation}
\delta \left( -\frac 12g_{\alpha \beta }\,dA^{{}\alpha }dA^{{}\beta
}-\lambda \,\left( \omega ^2d\ell _i^2-d\ell _f^2\right) \right) =0\,.
\label{var princ}
\end{equation}
We get 
\begin{equation}
dA^{{}\alpha }=\chi \Delta A^{{}\alpha }\quad \text{where}\quad \chi \equiv
\frac 1{1-\omega ^2+1/2\lambda }\,\,.  \label{main}
\end{equation}
The multiplier $\lambda $, or equivalently the quantity $\chi $, is
determined substituting back into the constraint (\ref{w constr}). From eq.(%
\ref{distances}) we get $d\ell _i=\chi \Delta \ell $ and $d\ell _f=(1-\chi
)\Delta \ell $, and therefore \cite{footnote3} 
\begin{equation}
\chi =\frac 1{1+\omega }\quad \text{and}\quad \lambda =\frac 1{2\omega
(1+\omega )}\,.
\end{equation}
Thus, the intermediate state $A_{{}\omega }$ selected by the maximum entropy
method is such that 
\begin{equation}
d\ell _i+d\ell _f=\Delta \ell \,.
\end{equation}
The geometrical interpretation is obvious: the triangle defined by the
points $A_i$, $A_{{}\omega }$, and $A_f$ (fig.1) degenerates into a straight
line. This is sufficient to determine a short segment of the trajectory:\
all intermediate states lie on the straight line between $A_i$ and $A_f$.
The generalization beyond short trajectories is immediate: if any three
nearby points along a curve lie on a straight line the curve is a geodesic.
Note that this result is independent of the value $\omega $ so the potential
consistency problem we had identified earlier does not arise.

To summarize, the answer to our question Q2 is simple and elegant:

\begin{description}
\item[ED]  The expected trajectory is the geodesic that passes through the
given initial and final states.
\end{description}

\noindent This is the main result of this paper. As promised, in entropic
dynamics the motion is predicted on the basis of a `principle of inference',
the principle of maximum entropy, and not from a `principle of physics'.

The dynamics ED was derived in an unusual way and one should expect some
unusual features. Indeed, they become evident as soon as one asks any
question involving time. For example, ED determines the vector tangent to
the trajectory $dA^{{}\alpha }/d\ell $, but not the actual `velocity' $%
dA^{{}\alpha }/dt$. The reason is not hard to find: nowhere in question Q2
nor in any implicit background information is there any reference to an
external time $t$.

Additional information is required if one is to find a relation between the
distance $\ell $ along the trajectory and the external time $t$. In
conventional forms of dynamics this information is implicitly supplied by a
`principle of physics', by a Hamiltonian which fixes the evolution of a
system relative to external clocks. But Q2 makes no mention of any external
universe; the only clock available is the system itself, and our problem
becomes one of deciding how this clock should be read. We could, for
example, choose one of the variables $A^{{}\alpha }$, say $A^1$, as our
clock variable and arbitrarily call it intrinsic time. Ultimately it is best 
\emph{to define intrinsic time so that motion looks simple}.

A very natural definition consists in stipulating that the system moves with
unit velocity, then the intrinsic time $\tau $ is given by the distance $%
\ell $ itself, $d\tau =d\ell $. \emph{Intrinsic time is quantified change}.
A peculiar consequence of this definition is that intervals between events
along the trajectory are not a priori known, they are determined only after
the equations of motion are solved and the actual trajectory is determined.
This reminds us of the theory of General Relativity (GR).

An important feature of GR is the absence of references to an external time.
Given initial and final states, in this case the initial and final
three-dimensional geometries of space, the proper time interval along any
curve between them is only determined after solving the Einstein equations
of motion \cite{Baierlein62}. The absence of an external time has been a
serious impediment in understanding the classical theory \cite{York} --
because it is not clear which variables represent the true gravitational
degrees of freedom -- and also in formulating a quantum theory \cite
{Kuchar92} -- because of difficulties in defining equal-time commutators.

In the following section we rewrite the entropic dynamics in Lagrangian and
Hamiltonian forms and we point out some further formal similarities between
ED and GR.

\section{Formal developments}

The entropic dynamics can be derived from an `action' principle. Since the
trajectory is a geodesic, the `action' is the length itself, 
\begin{equation}
J[A]=\int_{\eta _i}^{\eta _f}d\eta \,L(A,\dot{A}),  \label{J[A]}
\end{equation}
where $\eta $ is an arbitrary parameter along the trajectory, the Lagrangian
is 
\begin{equation}
L(A,\dot{A})=\left( g_{\alpha \beta }\dot{A}^{{}\alpha }\dot{A}^{{}\beta
}\right) ^{1/2}\quad \text{and}\quad \dot{A}^{{}\alpha }=\frac{dA^{{}\alpha }%
}{d\eta }\,.  \label{Lagrangian}
\end{equation}

The action $J[A]$ is invariant under reparametrizations $A(\eta )\rightarrow
A(f(\eta ))$ provided the end points are left unchanged, $f(\eta _i)=\eta _i$
and $f(\eta _f)=\eta _f\,$. Indeed, when the transformation is
infinitesimal, $f(\eta )=\eta +\varepsilon (\eta )$, the corresponding
change in the action, 
\begin{equation}
\delta J=\left. \left( g_{\alpha \beta }\dot{A}^{{}\alpha }\dot{A}^{{}\beta
}\right) ^{1/2}\varepsilon (\eta )\right| _{\eta _i}^{\eta _f},
\end{equation}
vanishes provided $\varepsilon (\eta _i)=\varepsilon (\eta _f)=0$. As
pointed out in \cite{Teitelboim82} there is an important distinction between
the symmetries of a generally covariant theory such as GR and the internal
symmetries of a proper gauge theory. The action of a generally covariant
theory is invariant under those reparametrizations that are restricted to
map the boundary onto itself; for proper internal gauge transformations
there are no such restrictions. Thus ED shares with GR the fact that both
are generally covariant theories.

It is instructive to consider the analogous principle of least action for a
nonrelativistic particle. The standard Hamilton's principle requires
extremizing the action 
\begin{equation}
\int_{t_i}^{t_f}dt\left( \frac m2\delta _{ab}\frac{dx^{{}a}}{dt}\frac{%
dx^{{}b}}{dt}-V(x)\right) ,
\end{equation}
where $t$ is `physical' time, and the interval between initial and final
states $t_f-t_i$ is given. In contrast, Jacobi's principle of least action
for a particle with energy $E$ moving in a potential $V(x)$ determines the
trajectory by extremizing the action 
\begin{equation}
J[x]=\int_{\eta _i}^{\eta _f}d\eta \left( 2m\delta _{ab}\frac{dx^{{}a}}{%
d\eta }\frac{dx^{{}b}}{d\eta }\right) ^{1/2}\left( E-V(x)\right) ^{1/2}.
\label{Jacobi}
\end{equation}
There is no reference to any time $t$, the time interval between initial and
final states is not given, and the parameter $\eta $ is unphysical and
arbitrary. To determine the temporal evolution along the trajectory requires
an additional supplementary condition, 
\begin{equation}
\frac m2\delta _{ab}\frac{dx^{{}a}}{dt}\frac{dx^{{}b}}{dt}+V(x)=E\text{ }.
\end{equation}
Thus the ED action, eq.(\ref{J[A]}), is an action of the Jacobi type. The
natural choice for a supplementary condition that defines $\tau $ and
determines the evolution along the trajectory is 
\begin{equation}
g_{\alpha \beta }\frac{dA^{{}\alpha }}{d\tau }\frac{dA^{{}\beta }}{d\tau }%
=1\,\,.  \label{suppl cond}
\end{equation}
It is interesting that GR\ is also described by a Jacobi-type action \cite
{Brown89}. To explore this similarity further it is convenient to construct
the canonical (\emph{i.e.}, Hamiltonian as opposed to Lagrangian) version of
Jacobi's action.

The canonical momenta are given by 
\begin{equation}
\pi _{{}\alpha }=\frac{\partial L}{\partial \dot{A}^{{}\alpha }}=\frac{%
g_{\alpha \beta }\dot{A}^{{}\beta }}{\left( g_{\mu \nu }\dot{A}^{{}\mu }\dot{%
A}^{{}\nu }\right) ^{1/2}}\,  \label{momenta}
\end{equation}
and have unit magnitude, 
\begin{equation}
g^{{}\alpha \beta }\pi _{{}\alpha }\pi _{{}\beta }=1\,.  \label{mom mag}
\end{equation}
The canonical Hamiltonian vanishes identically, 
\begin{equation}
H_{\text{can}}(A,\pi )=\dot{A}^{{}\alpha }\pi _{{}\alpha }-L(A,\dot{A}%
)\equiv 0\,,
\end{equation}
because the Lagrangian is homogeneous of first degree in the $\dot{A}$'s.
Physically this is not surprising: the generator of time evolution can be
expected to vanish whenever there is no external time with respect to which
the system could possibly evolve. We are led to consider the canonical
action 
\begin{equation}
\int_{\eta _i}^{\eta _f}d\eta \left( \dot{A}^{{}\alpha }\pi _{{}\alpha }-H_{%
\text{can}}\right) =\int_{\eta _i}^{\eta _f}d\eta \,\dot{A}^{{}\alpha }\pi
_{{}\alpha }\,,
\end{equation}
but eq.(\ref{mom mag}) implies that unconstrained variations of the momenta $%
\pi _{{}\alpha }$ are not allowed. The correct variational principle
requires to extremize the action 
\begin{equation}
I[A,\pi ,N]=\int_{\eta _i}^{\eta _f}d\eta \,\left[ \dot{A}^{{}\alpha }\pi
_{{}\alpha }-N\,h(A,\pi )\right]
\end{equation}
where 
\begin{equation}
h(A,\pi )=\frac 12g^{{}\alpha \beta }\pi _{{}\alpha }\pi _{{}\beta }-\frac 12
\label{h}
\end{equation}
and $N(\eta )$ are Lagrange multipliers that enforce the constraint 
\begin{equation}
h(A,\pi )=0\,.  \label{h constraint}
\end{equation}
for each value of $\eta $. The overall factor of $1/2$ in eq.(\ref{h}) is
introduced for later convenience; it amounts to rescaling $N$. Variation of $%
I[A,\pi ,N]$ with respect to $A$, $\pi $, and $N$ yields the equations of
motion, 
\begin{equation}
\dot{\pi}_{{}\alpha }=-N\frac{\partial h}{\partial A^{{}\alpha }}\,,\quad 
\dot{A}^{{}\alpha }=N\frac{\partial h}{\partial \pi _{{}\alpha }}\,,
\label{eq motion}
\end{equation}
and eq.(\ref{h constraint}). Naturally there is no equation of motion for $N$
and it must be determined from the constraint. We obtain, 
\begin{equation}
N=\left( g_{\alpha \beta }\dot{A}^{{}\alpha }\dot{A}^{{}\beta }\right)
^{1/2},
\end{equation}
which, using the supplementary condition eq.(\ref{suppl cond}), implies 
\begin{equation}
d\tau =N\,d\eta \,.
\end{equation}
The analogue of $N$ in GR is called the lapse function, it gives the
increase of `intrinsic' time per unit increase of the unphysical parameter $%
\eta $. In terms of $\tau $ the equations of motion become 
\begin{equation}
\frac{d\pi _{{}\alpha }}{d\tau }=-\frac{\partial h}{\partial A^{{}\alpha }}%
\,\quad \text{and}\quad \frac{dA^{{}\alpha }}{d\tau }=\frac{\partial h}{%
\partial \pi _{{}\alpha }}\,\,.
\end{equation}
In reparametrization invariant or generally covariant theories there is no
canonical Hamiltonian (it vanishes identically) but there are constraints.
It is the constraints that play the role of generators of evolution, of
change. Accordingly, the analogue of eq.(\ref{h constraint}) in GR is called
the Hamiltonian constraint.

\section{Final remarks}

I have provided an answer to the question `Given the initial and final
states, what is the trajectory followed by the system?' The answer follows
from established principles of inference without invoking additional
`physical' postulates. The entropic dynamics thus derived turns out to be
formally similar to other generally covariant theories: the dynamics is
reversible; the trajectories are geodesics; the system supplies its own
notion of an intrinsic time; the motion can be derived from a variational
principle that turns out to be of the form of Jacobi's action principle
rather than the more familiar principle of Hamilton; and the canonical
Hamiltonian formulation is an example of a dynamics driven by constraints.

To conclude one should point out that a reasonable physical theory must
satisfy two key requirements: the first is that it must provide us with a
set of mathematical models, the second is that the theory must identify real
physical systems to which the models might possibly apply. The entropic
dynamics we propose in this paper satisfies the first requirement, but so
far it fails with respect to the second; it may be a reasonable theory but
it is not yet `physical'. There are formal similarities with the general
theory of relativity and one wonders: Is this a coincidence? Whether GR will
in the end turn out to be an example of ED is at this point no more than a
speculation. A more definite answer hinges on the still unsettled problem of
identifying those variables that describe the true degrees of freedom of the
gravitational field \cite{York}\cite{Kuchar92}.


\begin{thebibliography}{99}
\bibitem{footnote3}  There is a second solution to eq.(\ref{var princ}), $%
\chi =(1-\omega )^{-1}$. It corresponds to minimizing rather than maximizing 
$S$.
\end{thebibliography}
\end{document}